\documentclass[conference]{IEEEtran}
\IEEEoverridecommandlockouts
\usepackage{cite}
\usepackage{amsmath,amssymb,amsfonts}
\usepackage{algorithmic}
\usepackage{algorithm}
\usepackage{graphicx}
\usepackage[caption=false]{subfig}
\usepackage{textcomp}
\usepackage{xcolor}
\usepackage{color,soul}
\usepackage{booktabs}
\usepackage[inline]{enumitem}
\usepackage{siunitx}
\usepackage[all]{nowidow}
\def\BibTeX{{\rm B\kern-.05em{\sc i\kern-.025em b}\kern-.08em
    T\kern-.1667em\lower.7ex\hbox{E}\kern-.125emX}}
\usepackage[color]{changebar}
\usepackage{soul}
\usepackage{xspace}
\cbcolor{black}
\sethlcolor{yellow}

\usepackage[]{fancyhdr} %
\newcommand{\changefont}{\fontsize{9}{9}\selectfont}
\usepackage{url}
\begin{document}

\title{Reinforcement Learning--Based\\Transient Response Shaping for Microgrids \\

\thanks{{This work is supported in part by the National Science Foundation (NSF) under awards ECCS-1953198 and ECCS-1953213 and in part by the Commonwealth Cyber Initiative, an investment in the advancement of cyber R\&D, innovation, and workforce development (www.cyberinitiative.org), and in part by the U.S. Department of Energy's Office of Energy Efficiency and Renewable Energy (EERE) under the Solar Energy Technologies Office Award Number 38637. The views expressed herein do not necessarily represent the views of the U.S. Department of Energy or the United States Government.}}
}

\author{\IEEEauthorblockN{Ashwin Venkataramanan and Ali Mehrizi-Sani}
\IEEEauthorblockA{{The Bradley Department of Electrical and Computer Engineering} \\
{Virginia Polytechnic Institute and State University}\\
Blacksburg, VA 24061 \\
\{vashwin, mehrizi\}@vt.edu}
}
\maketitle
\thispagestyle{fancy}
\pagestyle{fancy}

\begin{abstract}
This work explores the usage of a supplementary controller for improving the transient performance of inverter--based resources (IBR) in microgrids. The supplementary controller is trained using a reinforcement learning (RL)--based algorithm to minimize transients in a power converter connected to a microgrid. The controller works autonomously to issue adaptive, intermediate set points based on the current state and trajectory of the observed or tracked variable. The ability of the designed controller to mitigate transients is verified on a medium voltage test system using PSCAD/EMTDC. 
\end{abstract}

\begin{IEEEkeywords}
Adaptive control, machine learning, reinforcement learning, set point modulation, transient response.
\end{IEEEkeywords}

\section{Introduction}
Improper tuning of control system parameters in inverter--based resources (IBR) could result in small-, large-signal instabilities, while also resulting in violation of operating limits of individual equipment~\cite{MG_stability_def},~\cite{MG_practical_stability}. On a given system, for a specific operating condition, several methods, including development of a small-signal dynamic model~\cite{katiraei}, could be utilized to accurately select control parameters to optimize the performance of IBRs. However, these factors often lead to significant deterioration of their performance:

\begin {itemize}

\item Changes in parameters due to microgrid reconfiguration. 

\item Replacement of synchronous machines or condensers in microgrids with IBRs resulting in reduced damping.

\item Plug-and-play operation resulting in occasional mismatch of controllers for a specific power system apparatus~\cite{NREL_roadmap}.

\item Dependency on IBRs to provide ancillary support~\cite{Delille} to the grid in case of faults or contingencies requiring drastic changes to setpoints of IBRs, which results in operation outside the design margins. 

\end {itemize}

While several methods are available in literature to overcome the above issues, they have the disadvantage of either being model-based or requiring dedicated communication links. Reference~\cite{SPAACE1} proposed set point automatic adjustment with correction-enabled (SPAACE), a strategy of scaling the set point issued to a power converter depending on the predicted value of the tracked variable. Compared to other methods, SPAACE is autonomous and model-free. Additionally, SPAACE is a black-box control method, that is, it does not require access to the internal parameters of the existing controller. Reference~\cite{SPAACE_theory} establishes the theoretical foundations for this strategy, and it is demonstrated through hardware experiments in~\cite{SPAACE_practical}. However, the previous strategies for set point adjustment or modulation utilize a fixed level of scaling for the set points. Reference~\cite{SPAACE_practical} recommends a scaling of 0.2 pu for a smooth response and for preventing unintended contingencies. While this results in noticeable improvements to both settling times and overshoots, there is scope to further improve the performance using adaptive scaling of set points, specifically, in a system where inverters are expected to perform grid-supporting functions. 

Data-driven methods could be utilized to make this strategy adaptive, while retaining its model-free characteristics. Specifically, reinforcement learning (RL) provides a suite of methods to learn from experiences towards a goal through trial and error. Reference~\cite{RL_review_1} summarizes the recent applications of RL in power systems. RL is utilized in~\cite{nikita} to improve primary frequency response in a microgrid by utilizing conservative voltage reduction (CVR); this paper utilizes RL to train a multi-layer perceptron in reducing the voltage of distribution systems to improve their dynamic frequency response. RL is also utilized in~\cite{Glavic} to improve transient stability in a multi-area system; in~\cite{Hadidi}, Q-learning is utilized to improve stability in wide-area systems. 

In this work (RL-SPAACE), RL is utilized to develop a supplementary controller that issues adaptive set points based on the current state and trajectory of the response variable in a single input, single output (SISO) control system of an IBR. In comparison to~\cite{SPAACE1,SPAACE_theory,SPAACE_practical}, the range of set point manipulations are also extended, thus providing the controller with a wider degree of freedom. RL-SPAACE retains the autonomous and model-free characteristic of SPAACE, while incorporating adaptiveness to the control logic. 

This paper is organized as follows. Section~\ref{sec:sec2} introduces RL-SPAACE, and establishes certain preliminaries for utilizing RL concepts in designing the supplementary controller. Section~\ref{sec:trainingstrategy} introduces the algorithm utilized for training the controller towards the desired goal, and section~\ref{sec:casestudies} tests the designed controller on a microgrid test system and verifies its transient performance. The final section concludes the paper. 

\section{Proposed RL-Based Controller}\label{sec:sec2}
\subsection{Introduction to RL-SPAACE}\label{sec:IntroRLS}

\begin{figure}[!htbp]
\centering
\includegraphics[width=3.4in, trim=10 10 10 10,clip]{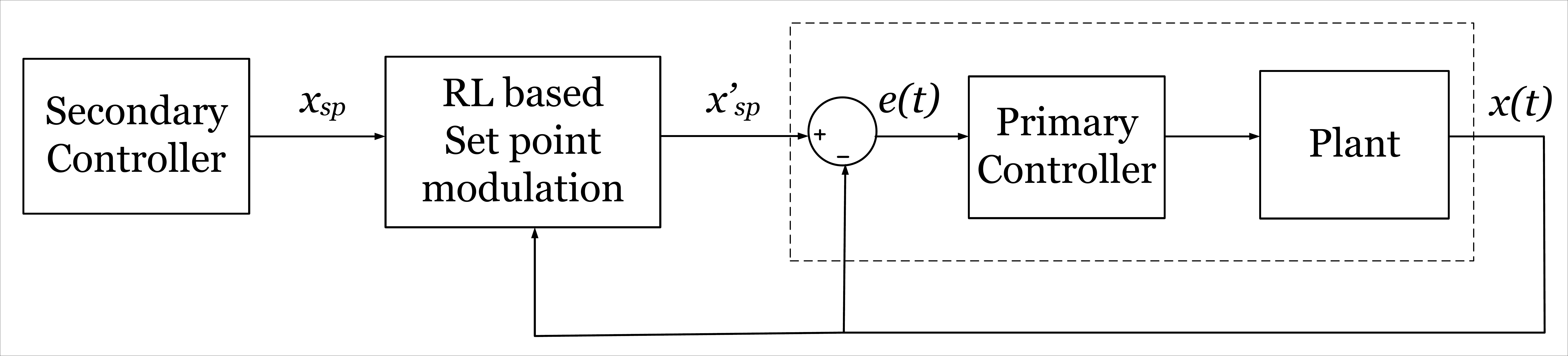}
\caption{IBR control scheme with the proposed supplementary controller.}
\label{fig:RLS_schematic}
\end{figure}

Fig.~\ref{fig:RLS_schematic} shows the schematic of the desired supplementary control action. RL-SPAACE operates utilizing signals outside the boundary of the power converter apparatus as shown in Fig.~\ref{fig:RLS_schematic}, thus enabling black-box control. Fig.~\ref{fig:RLS_schematic} shows RL-SPAACE issuing adjusted set points to a primary control process. The primary controller refers to any entity that utilizes the error from feedback to process it control signals. For an IBR, this could be a current, voltage, or power controller often utilized at the device level. The strategy is also applicable to nested control systems as long as the inner control loops are quicker than the successive stages. The secondary controller refers to the entity that issues slow set point changes to the primary control process. In a microgrid, this usually refers to a regional controller that issues desired real and reactive power set points to individual IBRs.

In RL-SPAACE and its previous versions, the actual set point ($x_{sp}(t)$) and the current value $(x(t))$ of the response variable are the only parameters utilized in deciding the modified set point ($x'_{sp}(t)$). While the previous versions utilize a fixed scaling factor $m(t)$, it is at the discretion of the controller in this work. 

The RL-Based controller has two primary objectives: \begin {enumerate*} [label=\itshape\alph*\upshape)]\item operate the IBR at the desired set point issued by the secondary controller in Fig.~\ref{fig:RLS_schematic} when the system is at steady state (at rest). \item when a transient occurs in the system, increase or decrease the set points in an adaptive manner to mitigate the transients. \end {enumerate*} Set point adjustments are issued as follows,
\begin{equation}
x'_{sp}(t)= 
\begin{cases}
\begin{aligned}

    &(1 + m(t))x_{sp}(t),     &\Delta x_{sp}(t)\geq0\;\text{and}\;|e(t)|>0\\
    &(1 - m(t))x_{sp}(t),     &\Delta x_{sp}(t)\leq0\;\text{and}\;|e(t)|>0\\
    &x_{sp}(t),            &\text{otherwise},

\label{eq:1}
\end{aligned}
\end{cases}
\end{equation}
 
In (\ref{eq:1}), $e(t)$ is the tracking error as shown in Fig.~\ref{fig:RLS_schematic}, and $\Delta x_{sp}(t)$ is the change in set point issued to the control system between two successive time steps. For a set point increase command, this value is less than zero, while it is greater than zero for a set point decrease command. As (1) suggests, RL-SPAACE utilizes a secondary logic to distinguish between positive, negative set point changes, and external disturbances. Depending on the nature of transient, the secondary logic activates a specific strategy of control using equation (1), and the concerned strategy remains active as long as $|e(t)|>0$.

\begin{figure}[!htbp]
\centering
\includegraphics[width=3.7in, trim=10 10 10 10,clip]{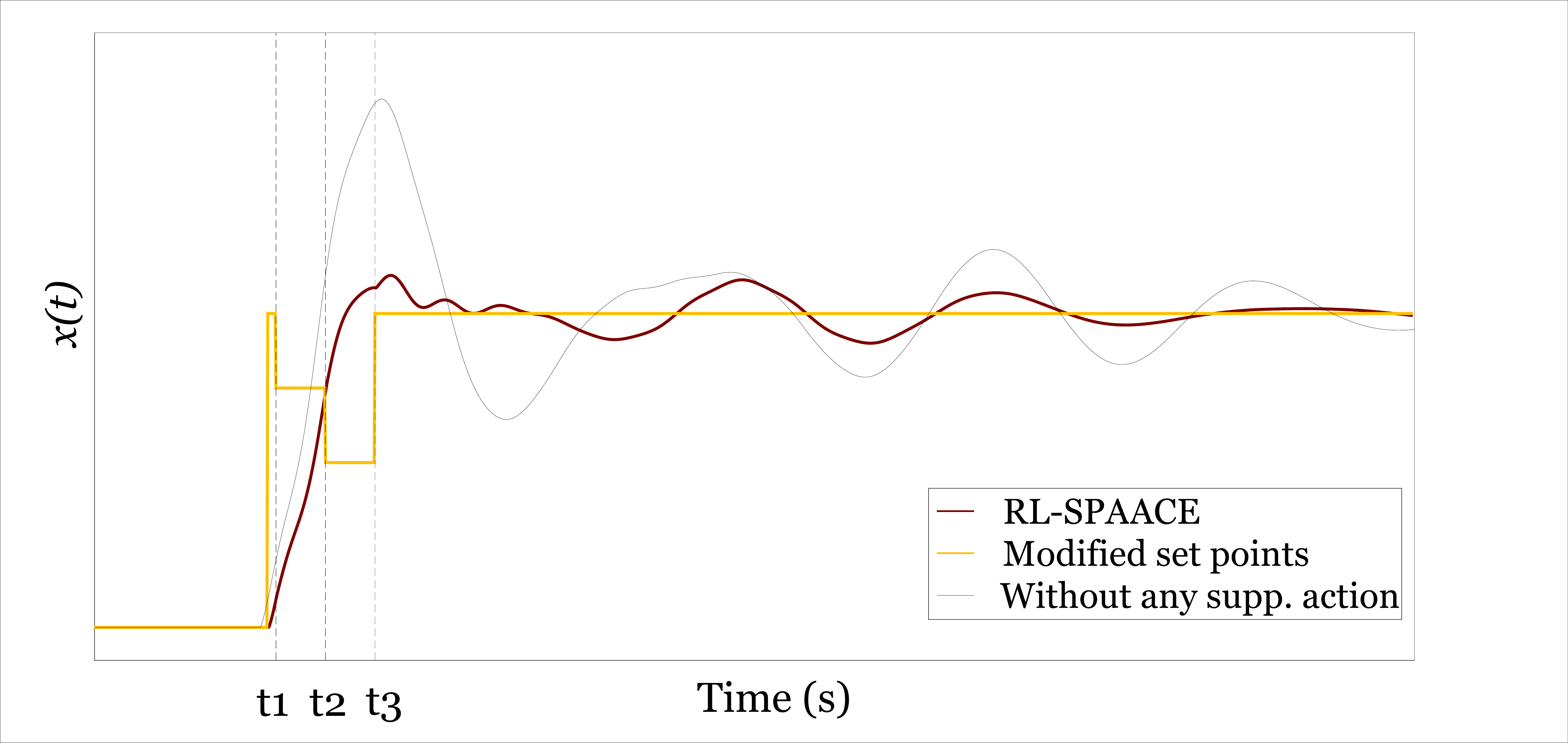}
\caption{Typical set point changes and response using RL-SPAACE.}
\label{fig:IntroRLS}
\end{figure}

Fig.~\ref{fig:IntroRLS} shows an example of the set point changes issued by RL-SPAACE to meet the desired objective. The example resembles a scenario where a negative set point change $\Delta x_{sp}(t)$ occurs, that is, a set point increase command from the secondary controller in a control system resembling Fig.~\ref{fig:RLS_schematic}. In Fig.~\ref{fig:IntroRLS}, at a time shortly before $t_1$, RL-SPAACE recognizes the set point change issued by secondary controller, and transmits this to the the primary controller and monitors the current value and trajectory of the response variable. Anticipating that the current trajectory will result in excessive overshoot, RL-SPAACE issues set point decrease commands at the times $t_1$ and $t_2$ before returning to the original set point at $t_3$.

\subsection{RL Formulation}\label{sec:probformulation}

In RL, the controller, also known as the agent, receives information on the current state $s_t$ of the system and performs an action $a_t$ on the environment. For each action, the agent receives a numerical reward $R_{t+1}$ and information on the next state of the system $s_{t+1}$. Based on the state $s_{t+1}$, the agent eventually selects a subsequent action $a_{t+1}$. Utilizing these inputs, the agent is trained to refine its sequence of actions, in other words, its control policy towards achieving a predetermined goal. The control policy learned by the controller is often stored in a look-up table or approximated to a function. While function approximation has found prominence in recent literature due to the development of other machine learning architectures, it does not provide an exact solution, and it is often useful in solving large problems. In this work, the tabular RL method is utilized as the method meets the accuracy levels required for controlling sensitive IBRs. 

The problem of optimal scaling of set points to achieve the desired trajectory of the response variable is a sequential decision-making problem in its essence, where the controller performs a sequence of actions over a period of time to meet its objective. The Markov decision process (MDP) provides a formal framework to represent these sequential decision making problems~\cite{sutton}. Once the problem is represented using MDP, RL could be utilized to compute the best possible solution. Reference~\cite{sutton} summarizes any goal-directed learning framework such as reinforcement learning using three signals---a measure of the current state of the system, a measure of the action performed by the agent on the system, and a measure of the impact of the actions performed by the agent. Accordingly, the elements that constitute an MDP are designed for the given problem as shown below,

The MDP is designed as follows:
\begin{itemize}
\item State $s$: $s$ is the discrete set of states representing the controlled system, also known as the environment. An obvious candidate to represent the state of the system is the controlled variable. However, as mentioned previously, the controller is designed to be active only during the dynamic state of the system. Utilizing the error $e$ into the control loop shown in Fig.~\ref{fig:RLS_schematic} provides a good measure of the system's dynamic state. Along with the error $e$, rate of change of error $\dot{e}$ is utilized to provide a better measure of the current state of system. Since MDPs require discrete state inputs, $e$ is discretized to 25 steps and $\dot{e}$ is discretized to 100 steps within the applicable range in each control paradigm determined by (1). 

\item Action $a$: $a$ is the finite set of actions performed by the controller or the agent on the environment. In this work, the action $a$ is the $m(t)$ chosen by the controller. $m(t)$ is discretized to five values within the range 0 to 0.95 when $\Delta x_{sp}<0$, and from 0 to 1.75 when $\Delta x_{sp}>0$. When $\Delta x_{sp}=0$, the value of $m(t)$ is discretized to seven values from $-$0.8 to 0.8.

\item Transition Probability $P$: The transition probability $P(s'|s,a)$ is the probability of transitioning from the state $s$ to the state $s'$ for an action $a$. Within the context of RL, the transition probability $P$ also refers to the model of a dynamic system in which the states exhibit the ``markov property." The model described here refers to the dynamic model computed by RL through interaction with the system, and it is not be confused with the physical model often utilized in control theory in designing controllers. The state input to the algorithm provides unique information about the state of the system at any point in time, and the system is fully observable. However, the state inputs are discretized to the controller, which occasionally results in stochastic behavior. 

\item Reward Function $R$: $R$ is the reward function which specifies the reward received by the agent once the system moves to a next state $s'$ as a result of an action $a$. The objective of the RL-Based controller is to improve the transient performance by minimizing overshoots and the settling times in the given system. In other words, this could be interpreted as driving the error towards zero as quickly as possible. This objective is incorporated into the reward function as follows,
\begin{equation}
r(s,a)= 
\begin{cases}
\begin{aligned}

    &-|e|- \lambda \Delta m, &|e| \leq 0.2 x_{\text{sp}} \;\\
    &-1000, &|e| > 0.2 x_{\text{sp}} \;

\label{eq:2}
\end{aligned}
\end{cases}
\end{equation}
In (2), $\lambda$ is the penalty factor associated with utilizing vastly different setpoints between two time steps. This is to encourage the controller to find an optimal solution without causing abrupt set point changes thus enhancing the quality of the response; $x_\text{sp}$ is the current set point issued by the secondary controller. In this work, $\lambda$ is chosen to be 0.1 for all training scenarios. From (2), it could be clearly deduced that the reward $r$ is directly obtained from the new system state $s'$. However, in a dynamic system that is stochastic, the state $s'$ is not unique for a specific action $a$ taken from the state $s$. As a result, (2) is utilized in computing a reward function $R$. This reward function is simply the average of all the rewards received by a particular state--action pair $(s,a)$.

\end{itemize}

The central idea of RL is to compute the value of state--action pairs for a given MDP. This value is the expected total payoff or return that can be obtained from a specific state or from a specific state for a selected action. As expected, for a sequential decision making process, this value function consists of two parts---an immediate and a discounted delayed reward. Utilizing the Bellman equation, the action-value function, $Q(s,a)$ is computed as follows for a particular state--action pair~\cite{sutton,Ernst},
\begin{equation}
Q(s,a)=R(s,a) + \gamma\sum_{s'} \bigg(P(s'|s,a)\max_{a'} Q({s'},{a'})\bigg)
\label{eq:3}
\end{equation}

In (\ref{eq:3}), $a'$ represents the set of available actions at the next state $s'$, and $\gamma$ is the factor by which the future or delayed rewards are discounted. $\gamma$ is required to be less than 1 for numerical stability. Equation~(\ref{eq:3}) can be recursively evaluated for each state--action pair to find its optimal value. Theoretically, each state--action pair under consideration has to be visited infinite times for the MDP solution to be optimal; however, a sufficient number of visits is expected to produce a solution that is near-optimal. 

\section{Training the Controller for the Desired Output}\label{sec:trainingstrategy}
The previous section established the preliminaries in setting up the problem for a RL-Based solution. The next step is to train the RL agent, RL-SPAACE, as efficiently as possible to meet the desired objectives. In any model-based RL algorithm, there are two essential processes---one is to utilize the experience from trial and error in computing the model of the environment, that is, building the transition probability matrix, $P$, the other is to ``plan," utilizing the learned model to search for better solutions. 

In Section~\ref{sec:probformulation}, the system was discretized to 2500 states. In each state, the agent was provided the freedom to choose any of the five to seven discrete control actions, resulting in a large $Q$-table or matrix. In traditional $Q$-value iteration methods, all the state--action pairs are updated over each sweep or iteration; the process is usually repeated over several iterations for the MDP to converge to a final solution. However, this is neither economical in terms of the usage of computational resources, nor in terms of the time spent in identifying the optimal solution. 

Several solutions are available in literature to improve the speed at which an optimal solution is identified with economical usage of computing resources. In this work, we utilize an algorithm known as \emph{real-time dynamic programming}~\cite{sutton}. In this algorithm, instead of updating $Q$ for all the state--action pairs in the system, $Q$ is updated only for the state--action pairs related to the new sample. This algorithm simply utilizes the trajectory of the system in updating the state--action pairs in the $Q$-table. Algorithm~1 in Fig.~\ref{fig:Algorithm1} lists the salient features of the training process. At step 5 in Algorithm~1, the agent, instead of updating the $Q$-table for all the state--action pairs, only the value corresponding to the recent state--action pair is updated. 

The algorithm also utilizes an $\epsilon$-greedy approach, where a random action is taken at $\epsilon\,\%$ of time instead of choosing the best action based on the $Q$ value. In this work, the $\epsilon$ value at the beginning is set to 0.3, which results in very high exploration. With a high degree of freedom available for the agent, a high value of exploration is necessitated at the beginning. Through the course of training, this value is reduced to 0.1. 

\begin{figure}[!t]
\begin{algorithm}[H]
\caption{RL-Based Set Point Modulation}
\begin{algorithmic}[1]
\FOR{$N$ episodes}
    \WHILE{time $\leq$ run time}
    \item $s\xrightarrow{}(e,\dot{e})$; Determine the discrete state $s$ of the system based on the current error and its rate of change. 
    \item $a\xrightarrow{}m(t)$; Discretize the action space $a$ to five or seven values depending on the training scenario.
    \item Using the index $\epsilon$, choose the best action according to the quality matrix, $Q$, or an exploratory action.
    \item Update the model $P(s'|s,a)$
    \IF{$i_d$ is within overshoot and undershoot limits}
    \item reward = $-|e_{t}|- \lambda \Delta m_{t}$
    \ELSE
    \item reward = $-1000$, high negative reward.
    \ENDIF
    \item Update the quality $Q(s,a)$ based on (3).
    \ENDWHILE
\ENDFOR    
\end{algorithmic}
\end{algorithm}
\caption{RL-SPAACE algorithm utilized in training the supplementary controller.}
\label{fig:Algorithm1}
\end{figure}

\begin{figure}[!t]
\centering
\includegraphics[width=3.55in, trim=10 10 5 10,clip]{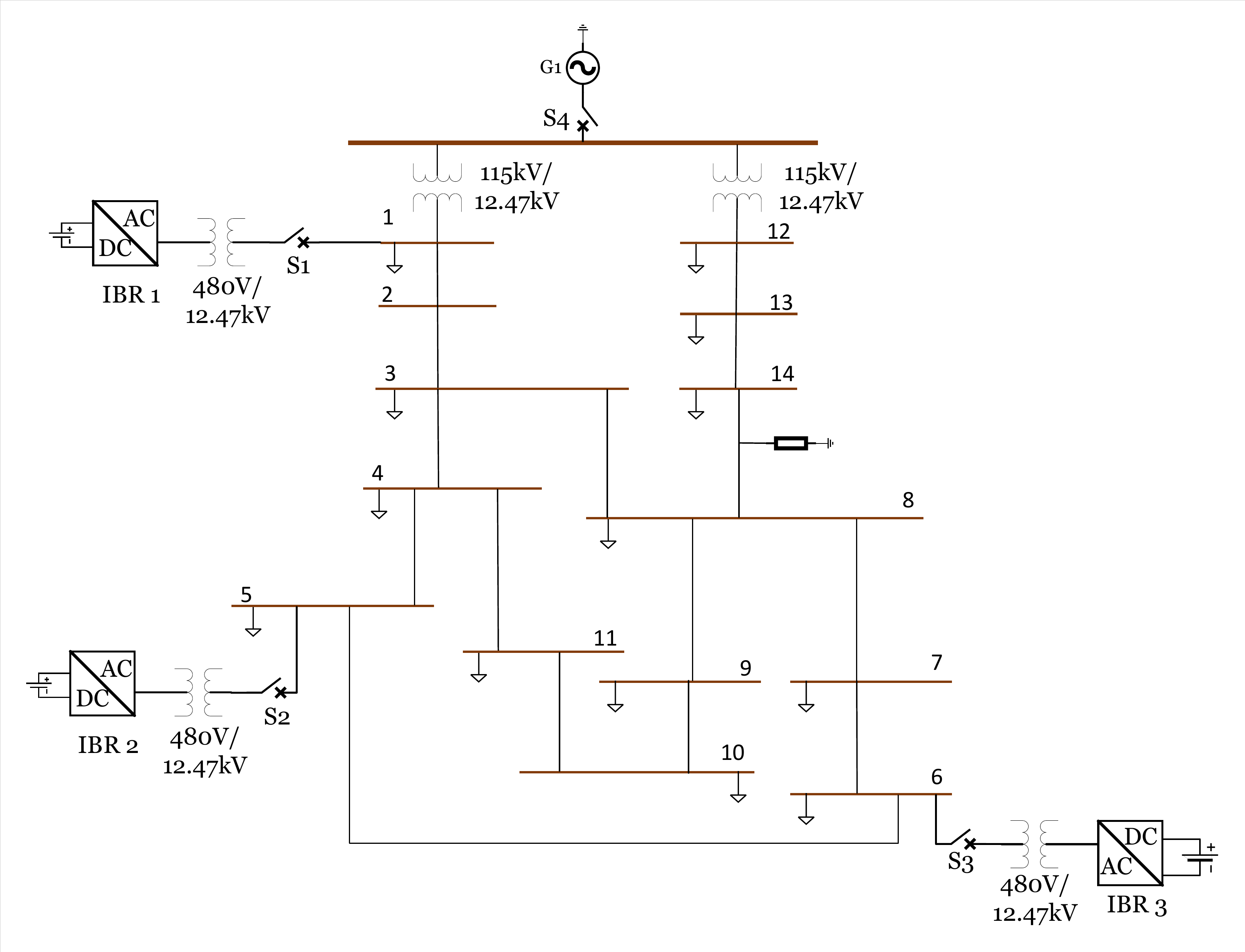}
\caption{CIGRE 14-bus North American medium-voltage test system with IBRs.}
\label{fig:CIGRE-14}
\end{figure}

The RL agent is programmed in python. The python based RL agent communicates with the PSCAD environment using the co-simulation toolbox available in PSCAD v5. The test system utilized in this work is the North American version of the CIGRE 14-bus medium voltage test system~\cite{CIGRE-14} shown in Fig.~\ref{fig:CIGRE-14}. In this test system, the rated voltage of the grid is \SI{115}{kV}, and the rated voltage of the downstream distribution feeders are \SI{12.47}{kV}. Two \SI{115}{kV}/\SI{12.47}{kV} step-down transformers connect the transmission grid to the distribution feeders. The IBRs in Fig.~\ref{fig:CIGRE-14} are integrated to the distribution feeder through a \SI{480}{V}/\SI{12.47}{kV} step-up transformer. The IBR connected to bus 1 is rated for \SI{6}{MVA}; the IBRs connected to buses 5 and 6 are rated for \SI{1}{MVA} and \SI{0.5}{MVA}, respectively. In each IBR, an average representation of the 3-phase voltage-sourced converter (VSC) is utilized. An ideal voltage source with an input voltage of \SI{1.2}{kV} is utilized to model the source behind the inverters; each inverter connects to the grid through an LCL filter. The switches~S1, S2, and S3 add or remove IBRs from the test system, while the switch~S4 isolates the test system from the grid.

The RL agent is initially trained to minimize transients when the IBRs operate in grid-following mode. The dominant portion of this work considers a power converter operating in grid-following mode. In order to meet the objectives mentioned in Section{~\ref{sec:IntroRLS}}, the controller is primarily trained to respond to set point changes from the secondary controller, and to operate in a resilient manner to any unplanned contingencies at its terminals or nearby buses. The value of the primary PI current controller coefficients at IBR1 and IBR2 are reduced intentionally to the following values: $k_P$=0.1 and $k_I$=0.001. This is done to mimic a mismatched controller or a lightly-damped system. The PI controller coefficient in IBR3 is set to $k_P$=0.18 and $k_I$=0.0015. The RL agent is trained using Algorithm~1 in Fig.~\ref{fig:Algorithm1}. 

In the first scenario, the RL agent is trained to limit transients when set point changes are issued to individual IBRs in Fig.~\ref{fig:CIGRE-14}. In another scenario, the RL agent is trained to limit transients when simultaneous set point changes are issued to IBRs at nearby buses. For this scenario, each IBR is equipped with an individual supplementary controller and it is allowed to simultaneously learn an independent control policy that is effective for local conditions. In both the scenarios, the system starts with the concerned IBR operating at the following set points: $i_d$ = 0 pu and $i_q$ = 0 pu. Once the system reaches steady state, a large set point change to $i_d$ with an upper limit of 1.1 pu is applied, while $i_q$ is set to 0 pu. The set point change to $i_d$ is determined at random in each iteration to cover a wide range of dynamics. Similarly, the RL agent is also trained to minimize transients when set point reductions are issued by the secondary controller. For this case, the value of $i_d$ is dropped from 1.0 pu to a value that is selected in random from 1.0 pu to 0.1 pu. In this scenario, $i_q$ is left unchanged at 0 pu. 

RL-SPAACE is also trained to enhance the resiliency of IBRs for unexpected contingencies. To train the RL agent for these situations, additional training scenarios are utilized. The RL agent is trained to minimize transients in IBR current, while operating in grid-following mode, for load energization events at their respective medium voltage bus. In each iteration, the IBR operates with the $i_d$ and $i_q$ set points of 1.0 pu and 0 pu, respectively, while a resistive load that is randomly selected between \SI{1}{MW} and \SI{10}{MW} is energized. 

Similarly, the RL agent is trained to minimize transients in IBR current when a symmetrical fault occurs at a medium voltage bus in the test system. Similar to the previous training scenario, the IBR operates with the $i_d$ and $i_q$ set points of 1.0 pu and 0 pu, respectively, while a 3-phase fault occurs at the respective medium voltage bus. The fault resistance is chosen randomly between \SI{0.01}{\ohm} and \SI{10}{\ohm}.

The test system in PSCAD communicates to the RL agent every \SI{100}{\micro\s} while training for set point adjustments. For other scenarios, the RL agent communicates every \SI{50}{\micro\s}. The RL agent begins training as soon as a the system enters the dynamic region ($e > 0$); the training ends for each episode when the terminal state is reached. A terminal state is one of the following: the value of error $e$ reaching zero after a transient, that is, the system returns to steady state, the value of error $e$ becoming higher than $0.2 x_\text{sp}$, and the end of simulation.

\begin{figure}
\centering
\includegraphics[width=3.7in, trim=6 10 10 10,clip]{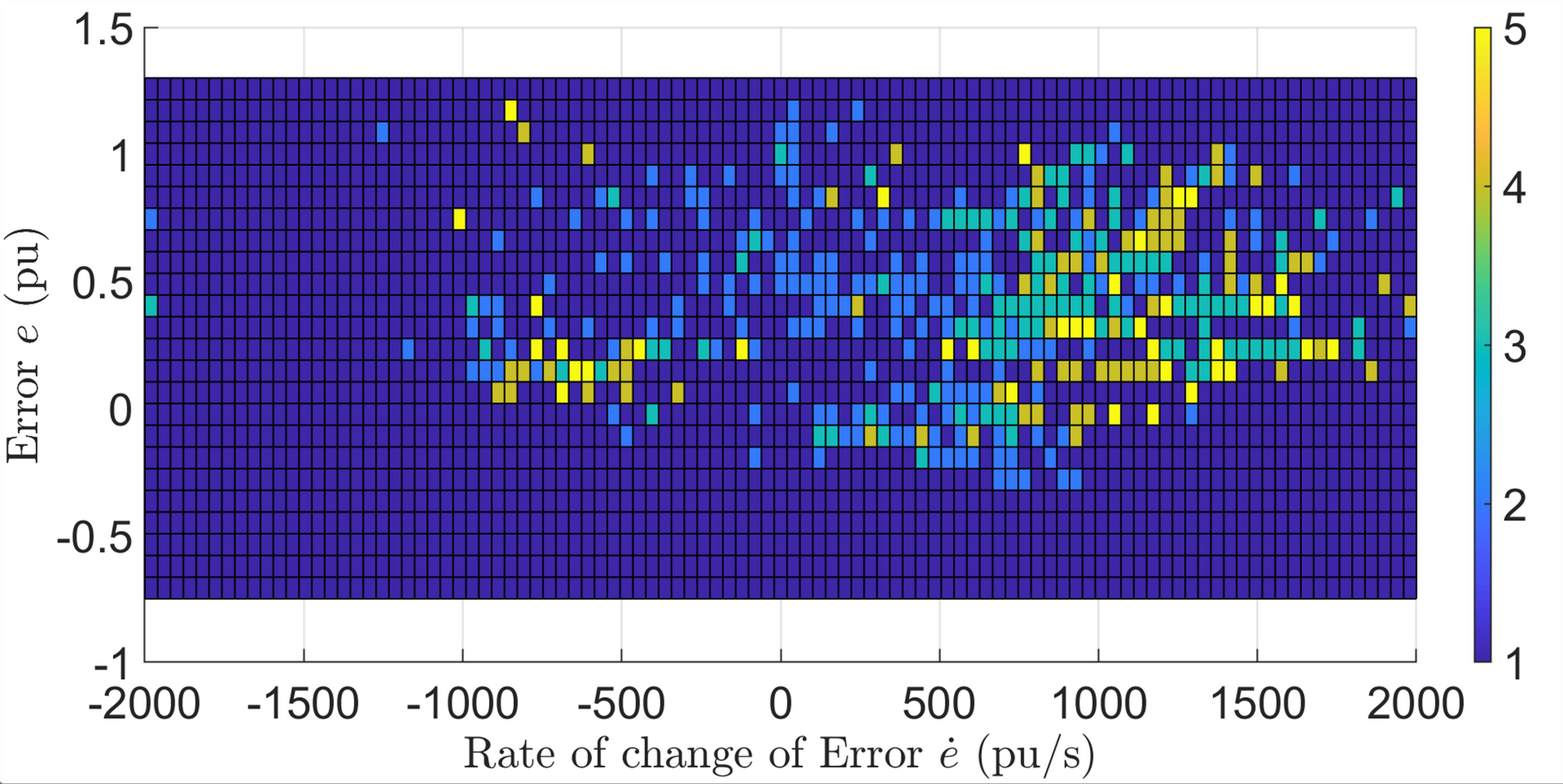}
\caption{An example decision matrix at the conclusion of training for a set point increase.}
\label{fig:decision_matrix}
\end{figure}

The training is concluded when the when the $Q$ converges to a final solution for the concerned state--action pairs. This is identified by plotting a decision matrix and monitoring for changes in this matrix for successive iterations. An example decision matrix is shown in Fig.~\ref{fig:decision_matrix}. This figure essentially represents the control policy learned by the RL agent. Each rectangle in the figure represents a specific system state, and the color on each rectangle specifies the action taken the agent at that state as denoted by the color bar on the right in the figure. 

\section{Case Studies}\label{sec:casestudies}
In this section, the performance of RL-SPAACE in minimizing transients is evaluated. The evaluations are performed using PSCAD/EMTDC with the RL-SPAACE controller programmed in python. Similar to the training setup, the test setup utilizes the co-simulation environment in PSCAD to establish communication between the test system and the trained controller. The performance is assessed on the test system in Fig.~\ref{fig:CIGRE-14}. 

The performance of RL-SPAACE is compared with a previous implementation of SPAACE with a linear predictor and with the performance of the primary controller without any supplementary control. SPAACE is also programmed in python, and the co-simulation toolbox in PSCAD establishes the communication with the controller. SPAACE utilizes the following strategy,

\begin{equation}
x'_{sp}(t)= 
\begin{cases}
\begin{aligned}

    &x_{sp}(t) (1 + m),     &x(t+t_{pred}) < x_{min}\\
    &x_{sp}(t) (1 - m),     &x(t+t_{pred}) > x_{max}\\
    &x_{sp}(t),            &\text{otherwise},

\label{eq:4}
\end{aligned}
\end{cases}
\end{equation}

The scaling factor $m$ and the prediction horizon $t_{pred}$ are adjusted in each case to obtain the best possible response. The maximum value of response variable $x_{max}$ and its minimum value $x_{min}$ are chosen depending on the individual device limits or ratings. 

\subsection{Case 1: Response to Current Set Point Changes}
\begin{figure}[!t]
    \centering
    \subfloat[]{\includegraphics[width=0.85\columnwidth, trim=8 8 6 8,clip]{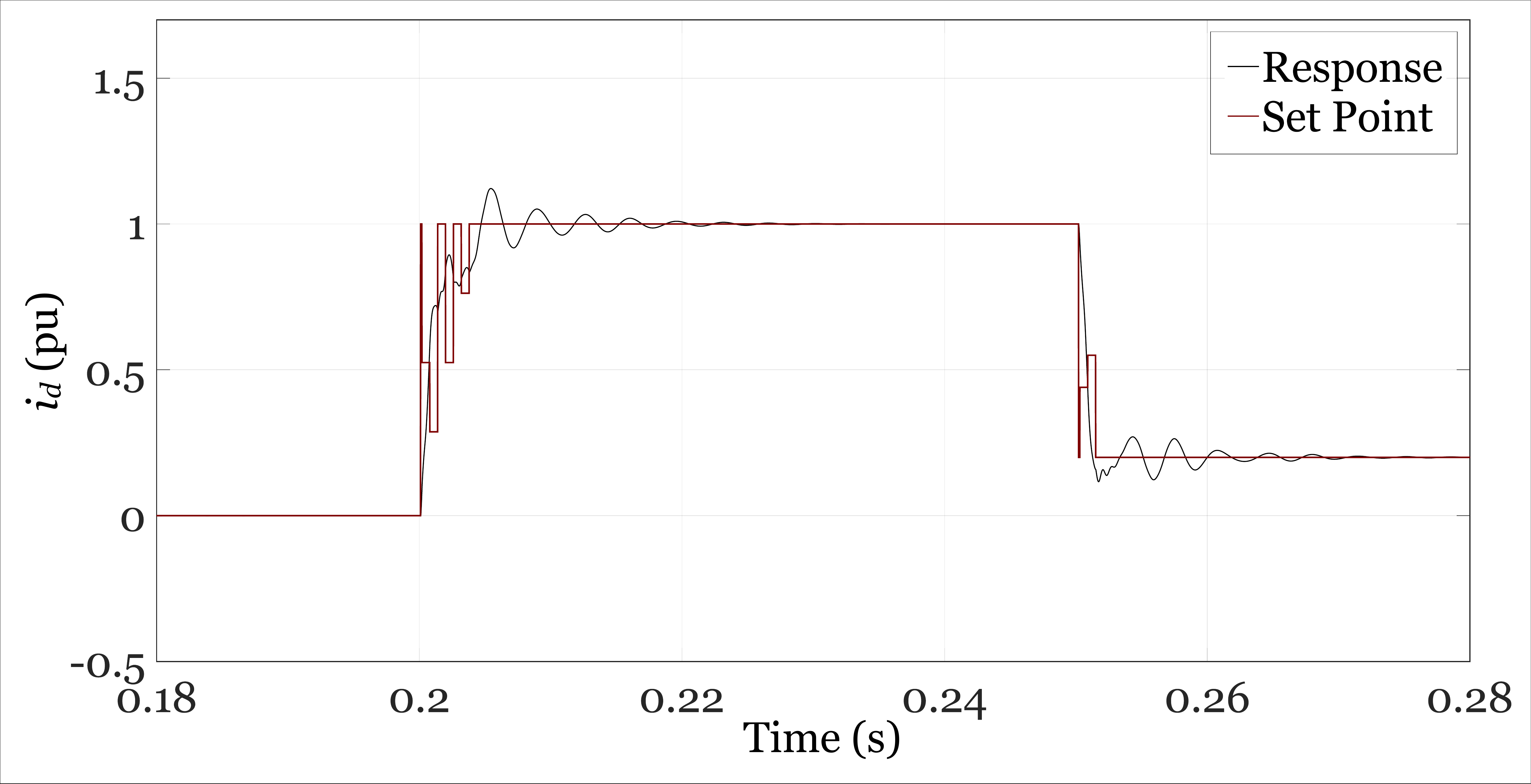}\label{fig:result1a}}\\[-0.2ex]
    \subfloat[]{\includegraphics[width=0.85\columnwidth, trim=8 8 6 8,clip]{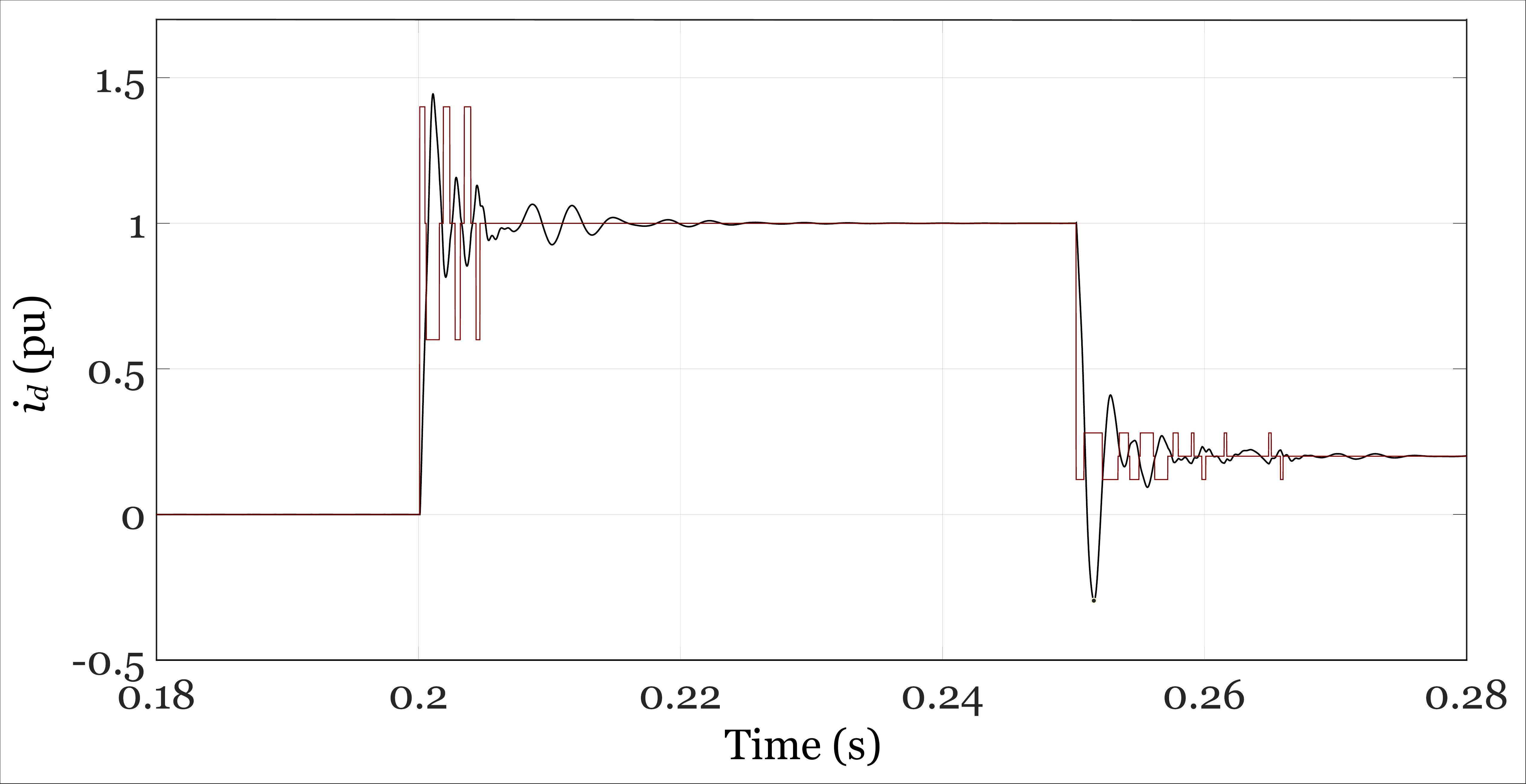}\label{fig:result1b}}\\[-0.2ex]
    \subfloat[]{\includegraphics[width=0.85\columnwidth, trim=8 8 6 8,clip]{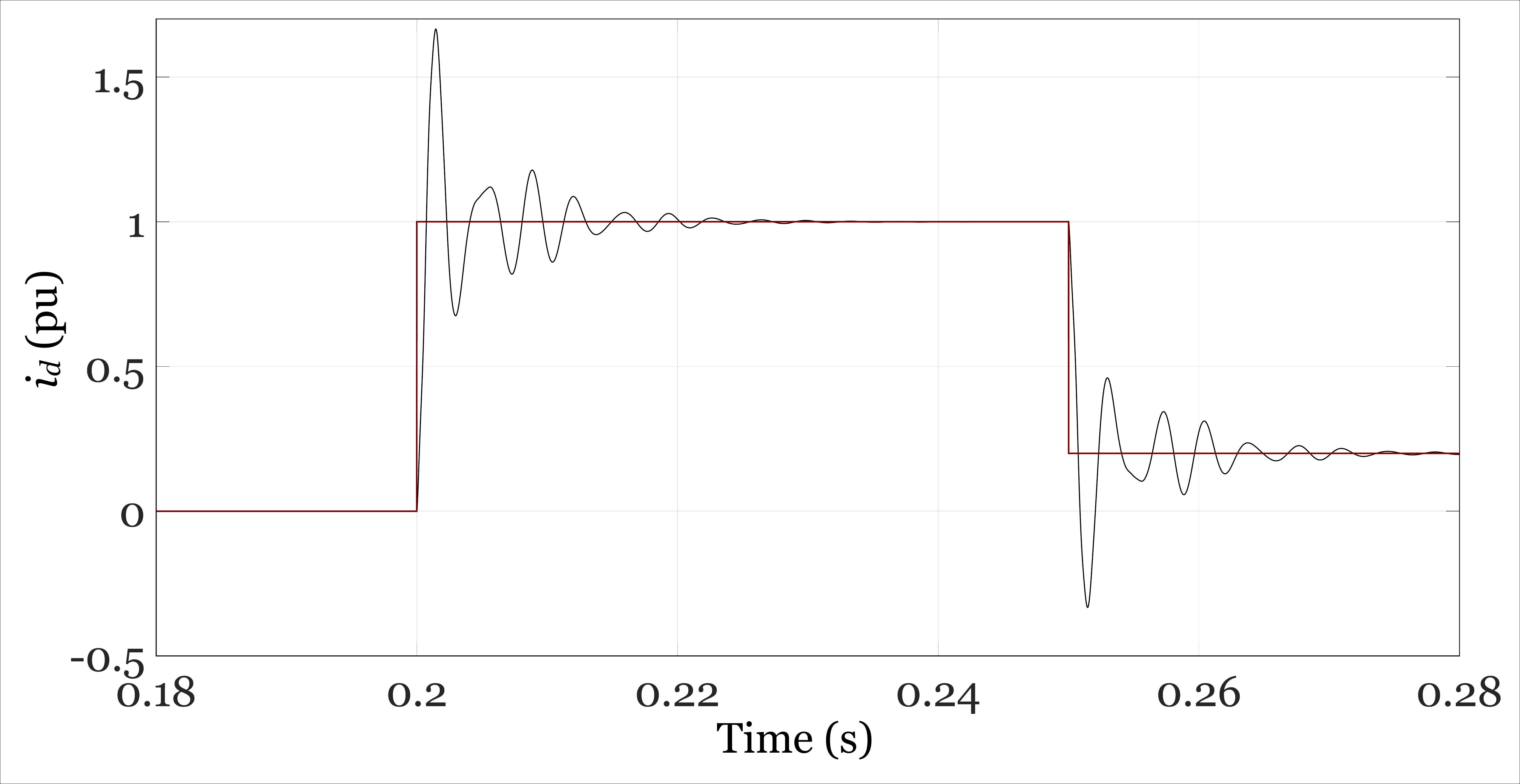}\label{fig:result1c}}
\caption{Case 1: Step change in IBR1 current from 0.0 pu to 1.0 pu (a) with RL-SPAACE, (b) with SPAACE, and (c) without any supplementary action.}
\label{fig:result1}
\end{figure}
This case study evaluates the performance of RL-SPAACE for a sudden, drastic increase, followed by a sudden decrease in set point of the d-axis component of inverter line current ($i_d$). In a typical microgrid, the sudden increase or decrease in set points resemble a scenario where an IBR, usually an energy storage device, is abruptly requested by the secondary controller to change its real power output in response to a system contingency.   

For this case, the switches S1 and S4 are closed while S2 and S3 are left open. At the start of the simulation, the IBR1 in Fig.~\ref{fig:CIGRE-14} is left idle, with the $i_d$ and $i_q$ set points at 0.0 pu. At a time $t=\SI{0.2}{s}$, the set point of $i_d$ is increased in step to 1.0 pu, while $i_q$ is set to 0.0 pu. Once $i_d$ and $i_q$ reach steady state, the set point of $i_d$ is decrease in step to 0.2 pu at time $t=\SI{0.25}{s}$, while $i_q$ is left unchanged at 0.0 pu. Fig.~\ref{fig:result1}\subref{fig:result1a} shows the response of the IBR1 current with supplementary action from RL-SPAACE. Fig.~\ref{fig:result1}\subref{fig:result1a} also shows the set point modulations enacted by RL-SPAACE to control the transients in IBR1 d-axis current. Figs.~\ref{fig:result1}\subref{fig:result1b} and~\ref{fig:result1}\subref{fig:result1c} show the same response with supplementary action from SPAACE with linear predication and the response without any supplementary control, respectively. 

\begin{table}[h]
\centering
\caption{Performance Indices for Increase in Set Point}
\begin{tabular}{l p{1.6cm} p{1.6cm}}\toprule
Control Strategy & Peak Overshoot (\%) & Cumulative Error $S_e$\\
\midrule
RL-SPAACE & 12 & 8.19\\
SPAACE & 44.3 & 6.86\\
No Supplementary Action & 65.7 & 9.01\\\bottomrule
\end{tabular}
\label{tab_case1}
\end{table}

\begin{table}[h]
\centering
\caption{Performance Indices for Decrease in Set Point}
\begin{tabular}{l p{1.6cm} p{1.6cm}}\toprule
Control Strategy & Peak Undershoot (\%) & Cumulative Error $S_e$\\
\midrule
RL-SPAACE & 37 & 6.10\\
SPAACE & 248 & 7.23\\
No Supplementary Action & 265 & 7.21\\\bottomrule
\end{tabular}
\label{tab_case1b}
\end{table}

Both SPAACE and RL-SPAACE communicate with the system every \SI{100}{\micro\s}. Additionally, for SPAACE, $m$ is chosen as 0.4 pu and the $t_{pred}$ is set to \SI{400}{\micro\s} for the linear predictor. Additionally, the maximum and minimum values of d-axis current as set to 1.2 pu and 0.8 pu, respectively. The key inferences from the results in Fig.~\ref{fig:result1} are shown in the Tables~\ref{tab_case1} and~\ref{tab_case1b}. As mentioned in Section~\ref{sec:probformulation}, RL-SPAACE is designed to reduce the error to 0 as quickly as possible. Similar to~\cite{SPAACE_practical}, the L2-norm of the error values at each time step is utilized to measure the performance of RL-SPAACE with respect to the aforementioned objective. This metric is listed as the cumulative error $S_e$ in Tables~\ref{tab_case1} and Tables~\ref{tab_case1b}. 

The maximum overshoot in response with supplementary action from RL-SPAACE is over 30\% lower for increase in set point compared to other strategies. Similarly, over 200\% improvement in peak undershoot is reported for set point reductions when RL-SPAACE is utilized. Tables~\ref{tab_case1} and~\ref{tab_case1b} also shows that RL-SPAACE performs better than the case without any supplementary action in maintaining the error values close to 0. In (4), SPAACE is designed to temporarily increase the set point when the predicted response variable is below the minimum level. This helps in increasing the rise time of the response, and maintaining the values of response comparatively closer to the issued set point, while deteriorating the overshoot performance. The effect could be clearly seen in Table~\ref{tab_case1} where SPAACE performs better than RL-SPAACE in the cumulative error $S_e$ metric, while having significantly higher overshoot.

\begin{figure}[!t]
    \centering
    \subfloat[]{\includegraphics[width=0.85\columnwidth, trim=8 8 6 8,clip]{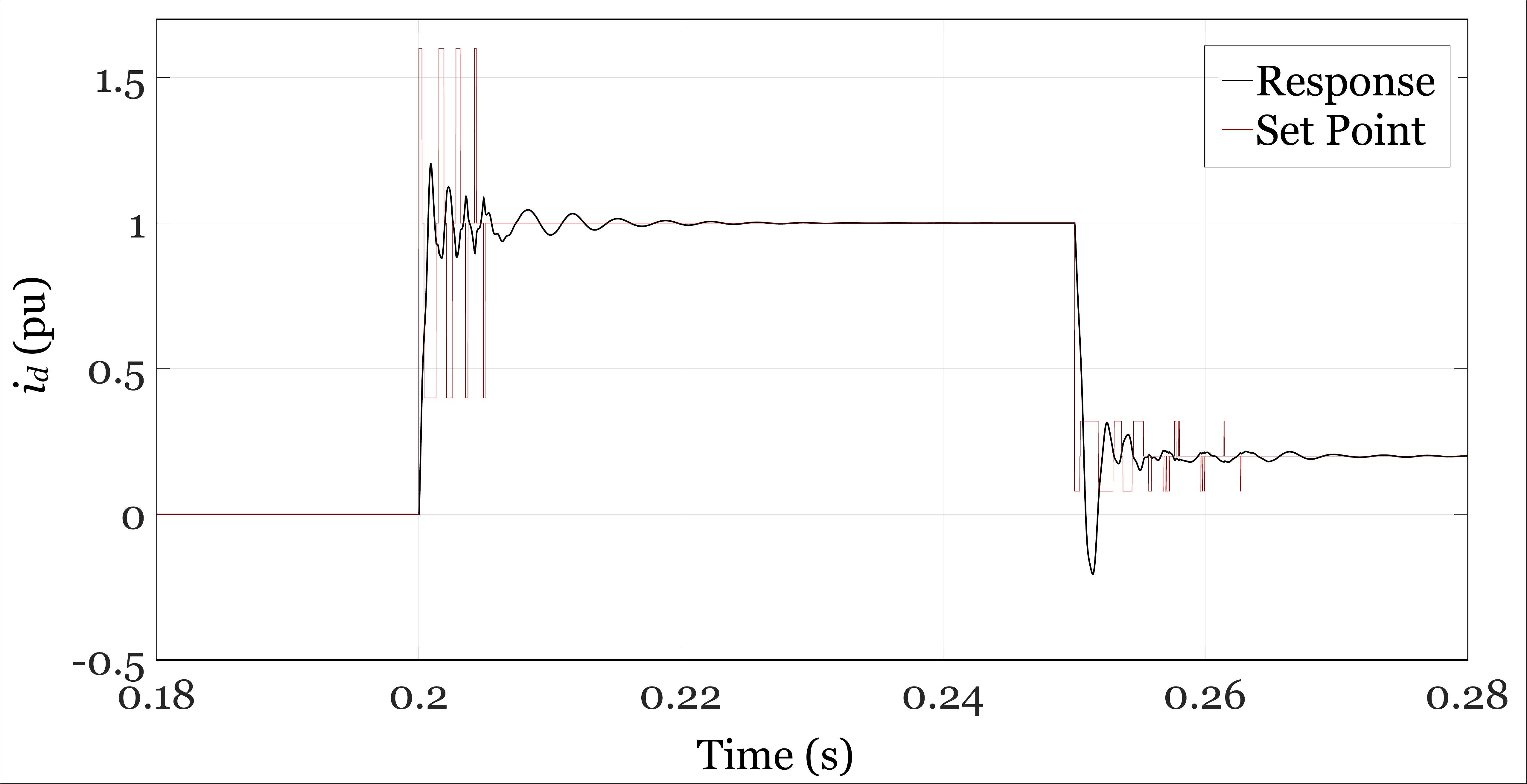}}\\[-0.2ex]
\caption{Case 1: Step change in IBR1 current from 0.0 pu to 1.0 pu with SPAACE utilizing higher communication rate.}
\label{fig:sub1}
\end{figure}

The performance of SPAACE could be improved by reducing the communication time between the controller and the system. For a communication time of \SI{10}{\micro\s}, with $m$ chosen as 0.6 pu and $t_{pred}$ set to \SI{800}{\micro\s}, the results could be improved as shown in Fig.~\ref{fig:sub1}. For the same set point changes as above, the peak overshoot is reduced to 20.3\% with a cumulative error 5.19 for the set point increase operation. This clearly shows the dependency of SPAACE on faster communication in systems with reduced damping levels. For set point decrease, the overshoot performance is similar to the previous study with higher communication time.

\subsection{Case 2: Response of Multiple IBRs to Simultaneous Set Point Changes}
In this case, the performance of RL-SPAACE when multiple IBRs are issued simultaneous set point changes is evaluated. In the 14-bus CIGRE test system shown in Fig.~\ref{fig:CIGRE-14}, switch S1 is opened, while the switches S2, S3, and S4 are closed. As mentioned in Section~\ref{sec:trainingstrategy}, the parameters of the inner current controller are set to $k_P$=0.1 and $k_I$=0.001 for IBR2. For IBR3, the same parameters are $k_P$=0.18 and $k_I$=0.0015. The values of the control coefficients are intentionally different to show that RL-SPAACE could be utilized in systems with different damping levels. 

\begin{figure}[!t]
    \centering
    \subfloat[]{\includegraphics[width=0.85\columnwidth, trim=8 8 6 8,clip]{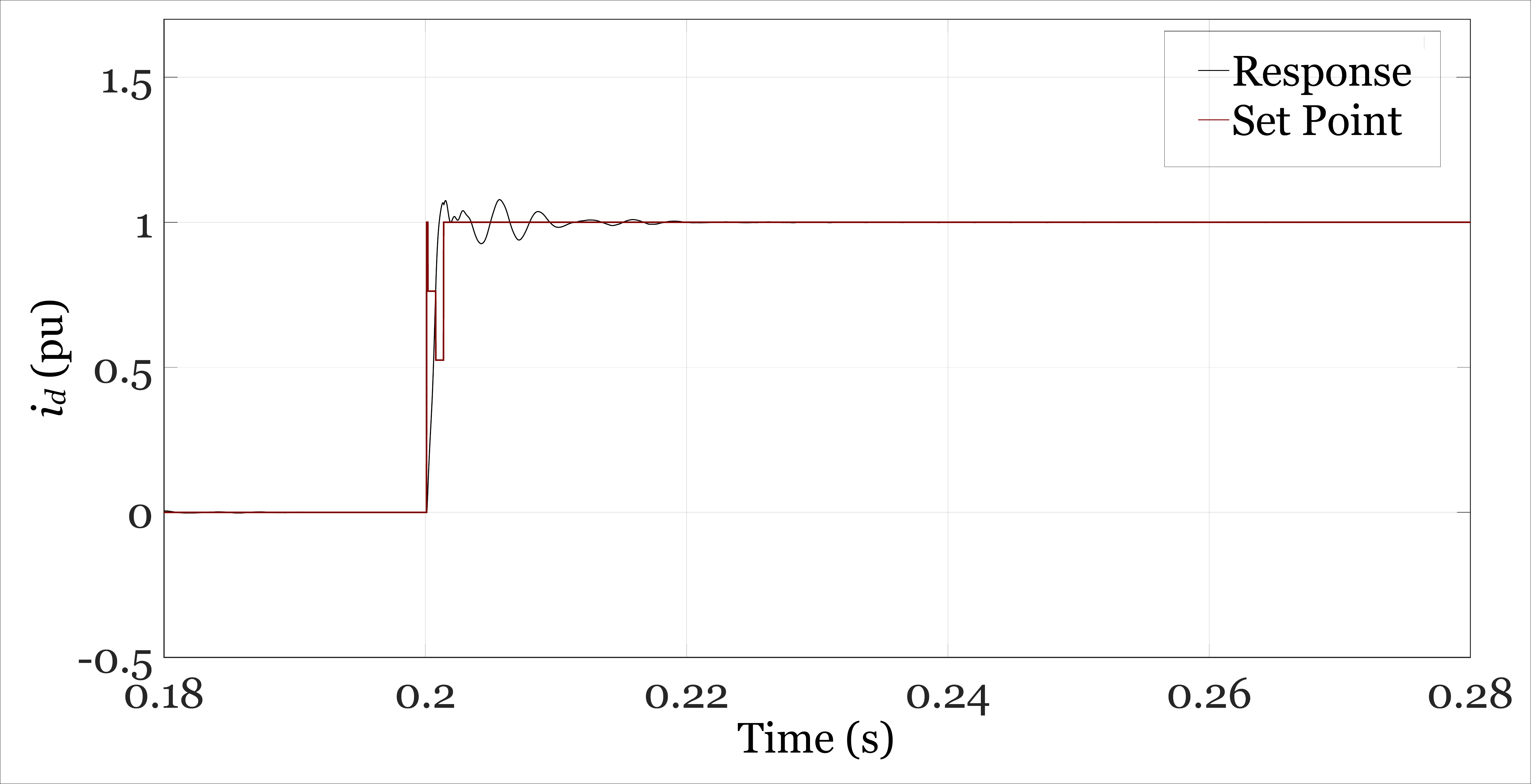}\label{fig:result2a}}\\[-0.2ex]
    \subfloat[]{\includegraphics[width=0.85\columnwidth, trim=8 8 6 8,clip]{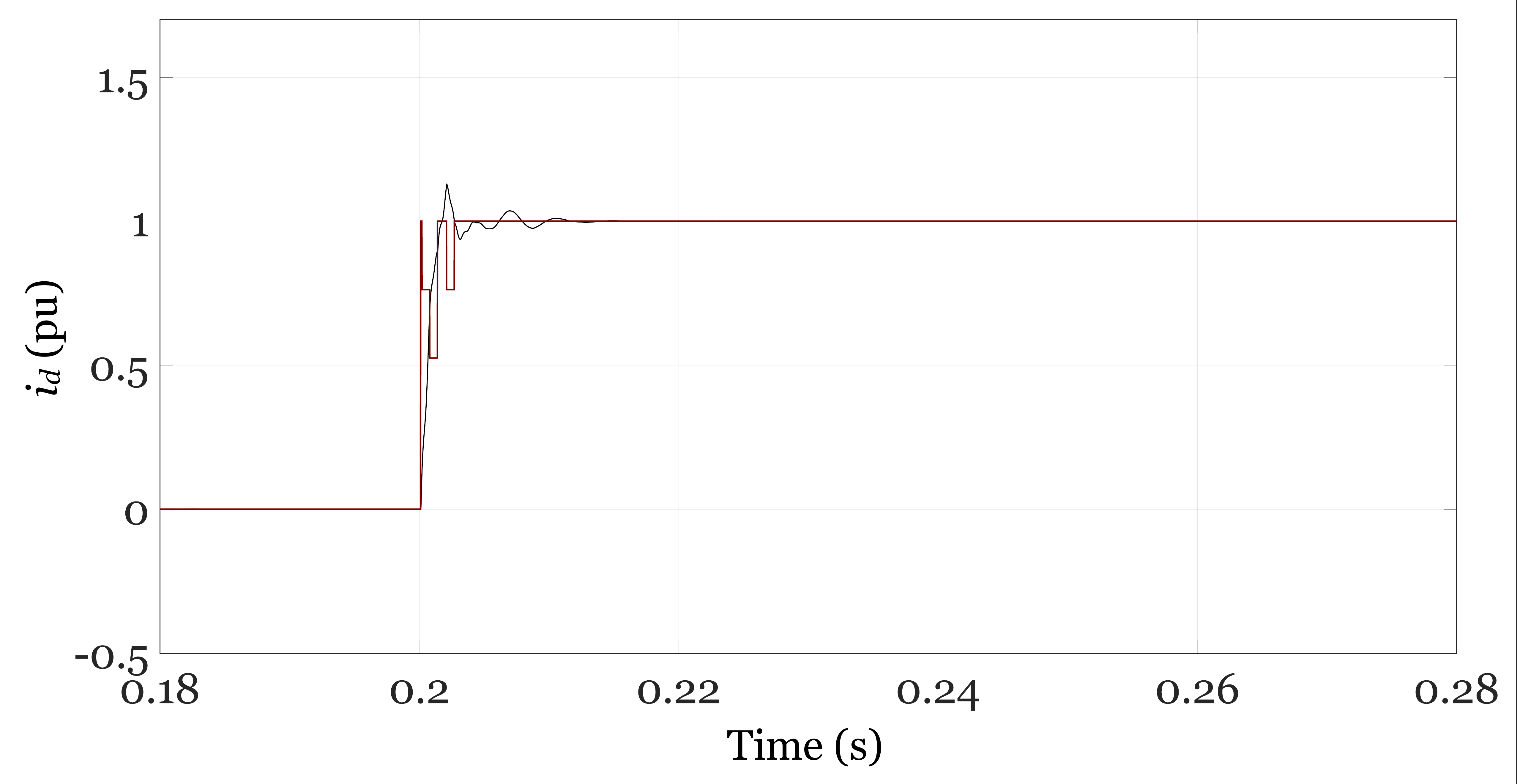}\label{fig:result2b}}\\[-0.2ex]
\caption{Case 2: Simultaneous set point change at (a) IBR2 and (b) IBR3 currents from 0.0 pu to 1.0 pu.}
\label{fig:result2}
\end{figure}

At the start of the simulation, IBR2 and IBR3 are left idle with the $i_d$ and $i_q$ set points at 0.0 pu. At a time $t=\SI{0.2}{s}$, the set point of $i_d$ is increase in step to 1.0 pu for both the IBRs; $i_q$ is left unchanged at 0.0 pu. Fig.~\ref{fig:result2}\subref{fig:result2a} shows the response of IBR2 with the set point adjustments from RL-SPAACE, while Fig.~\ref{fig:result2}\subref{fig:result2b} shows the response and set point adjustments for IBR~3. In Fig.~\ref{fig:result2}, the peak overshoot in IBR2 and IBR3 responses is 7.7\% and 12.5\%, respectively; the cumulative $S_e$ error in their responses is 6.75 and 6.89 respectively. As expected, the set point adjustments performed by RL-SPAACE at each IBR is unique suggesting that the control is local, depending on the requirements at the individual IBR.This case study illustrates that RL-SPAACE works as expected in a multi-IBR microgrid, where each IBR could be independently controlled utilizing an autonomous and decentralized controller. 

\subsection{Case 3: Response to Sudden Load Energization}
\begin{figure}[!t]
\centering
\includegraphics[width=\columnwidth, trim=10 10 10 10,clip]{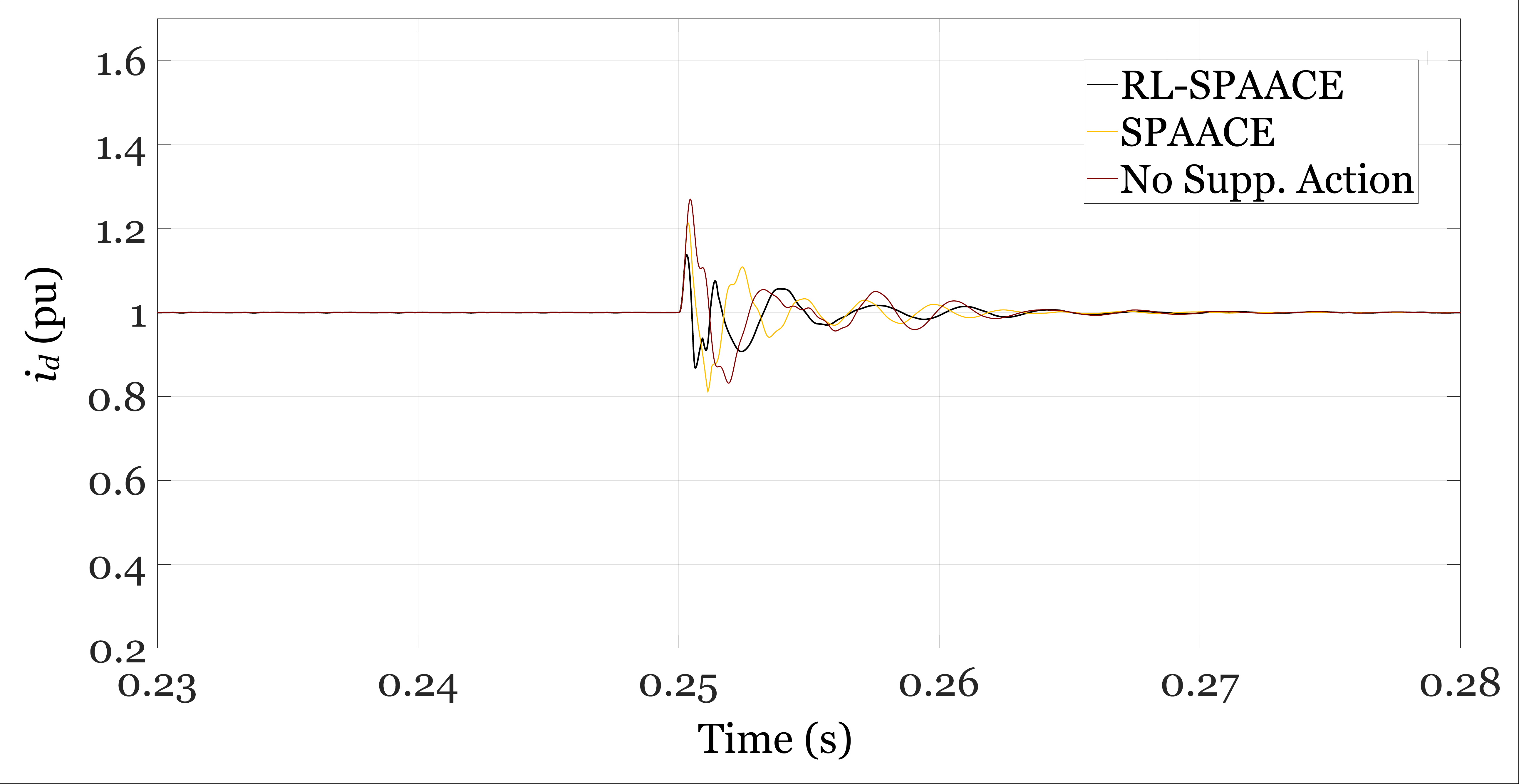}
\caption{Case 3: Response to sudden load energization at IBR2 terminals.}
\label{fig:Case3}
\end{figure}

The performance of RL-SPAACE in enhancing the resiliency of a microgrid is evaluated in this case. For this case, the switches S2 and S4 are closed and the switches S1 and S3 are left open in the test system in Fig.~\ref{fig:CIGRE-14}. At the start of the simulation, IBR2 is operated at its peak capacity with $i_d$ and $i_q$ set to 1.0 pu and 0.0 pu, respectively. At a time $t=\SI{0.25}{s}$, a \SI{10}{MW} resistive load connected to bus 5 is energized. 

\begin{table}[h]
\centering
\caption{Performance Indices for Case 3}
\begin{tabular}{l p{1.6cm} p{1.6cm}}\toprule
Control Strategy & Peak Overshoot (\%) & Cumulative Error\\
\midrule
RL-SPAACE & 13.9 & 1.46\\
SPAACE & 21.3 & 1.87\\
No Supplementary Action & 27.0 & 2.34\\\bottomrule
\end{tabular}
\label{tab_case2}
\end{table}

The ability of RL-SPAACE in mitigating transients in IBR2 current in response to the load energization is shown in Fig.~\ref{fig:Case3}.
Similar to previous cases, RL-SPAACE results in a significant reduction in peak overshoot compared to other methods. Table~\ref{tab_case2} also shows that RL-SPAACE performs significantly better in maintaining the tracking error close to 0. 
For this scenario, SPAACE utilizes a $m$ equal to 0.6 pu with $t_{pred}$ equal to \SI{100}{\micro\s}. Both SPAACE and RL-SPAACE communicate to the test system every \SI{50}{\micro\s}

\subsection{Case 4: Response to Symmetrical Faults}

\begin{figure}[h]
\centering
\includegraphics[width=\columnwidth, trim=10 10 10 10,clip]{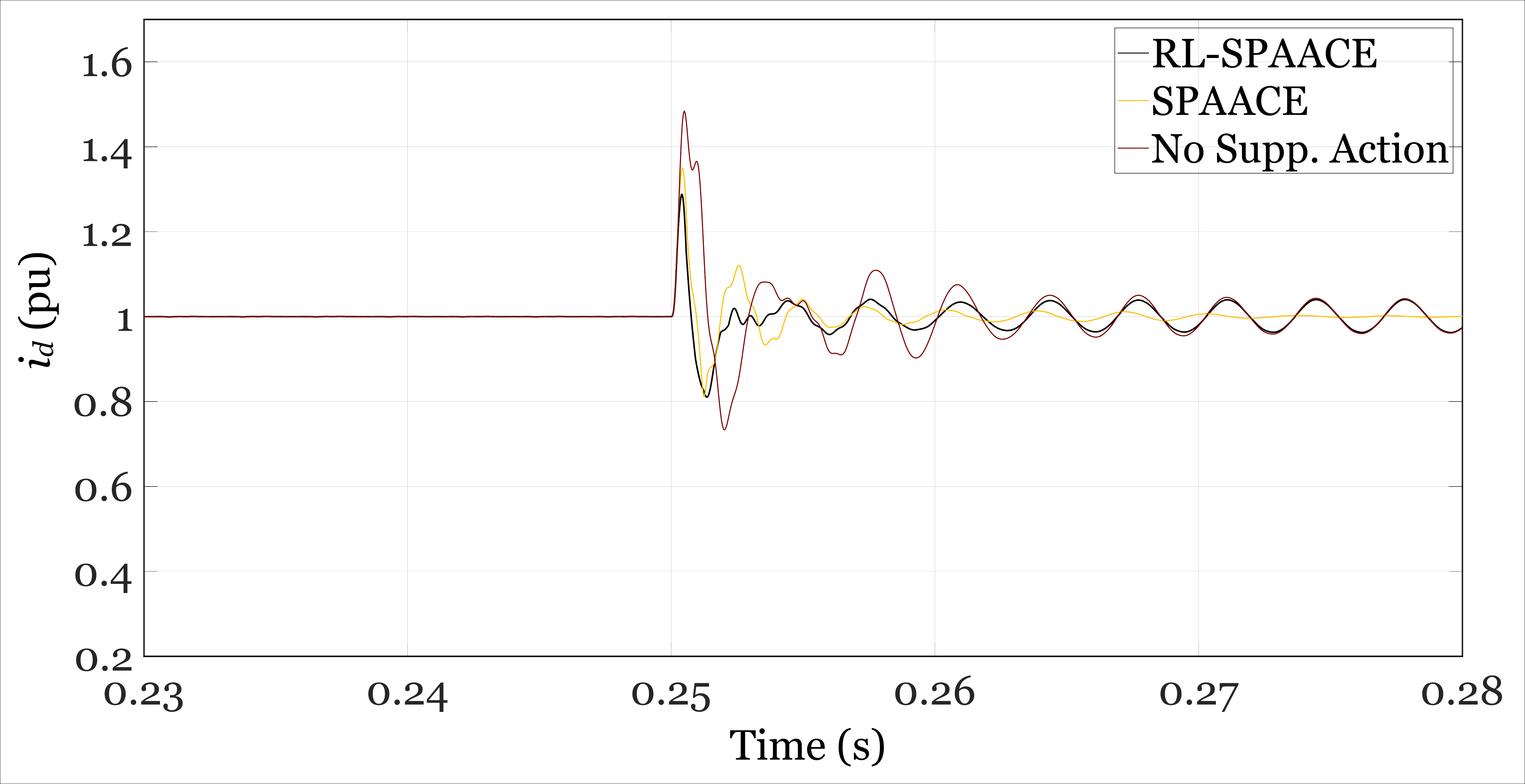}
\caption{Case 4: Response to a 3-phase fault at IBR2 terminals.}
\label{fig:Case4}
\end{figure}

Similar to Case 3, this case verifies the performance of RL-SPAACE in responding to a transient caused by a disturbance in the system. For this case, the topology of the test system from Case 3 is retained. At the start of the simulation, the IBR2 operates at its peak capacity with $i_d$ and $i_q$ set to 1.0 pu and 0.0 pu, respectively. At a time $t=\SI{0.25}{s}$, a 3-phase fault occurs at the bus 5 in the test system in Fig.~\ref{fig:CIGRE-14}. The fault resistance is set to \SI{0.01}{\ohm}. 

Fig.~\ref{fig:Case4} shows the results from this case. The results tabulated in Table~\ref{tab_case4} show that the peak overshoot with RL-SPAACE is reduced by almost 17\% with a significant reduction in cumulative error $S_e$ compared to the case without any supplementary action.

\begin{table}[h]
\centering
\caption{Performance Indices for Case 4}
\begin{tabular}{l p{1.6cm} p{1.6cm}}\toprule
Control Strategy & Peak Overshoot (\%) & Cumulative Error\\
\midrule
RL-SPAACE & 28.7 & 2.48\\
SPAACE & 34.9 & 2.49\\
No Supplementary Action & 48.3 & 4.84\\\bottomrule
\end{tabular}
\label{tab_case4}
\end{table}

\subsection{Case 5: Response to Voltage Set Point Changes}

The cases~1 to 4 evaluated the performance of the RL-SPAACE when the IBR is operated in grid-following or current control mode. In this case, RL-SPAACE is trained to minimize voltage transients in IBR terminal voltage when it operates in grid-forming mode. As mentioned in Section~\ref{sec:IntroRLS}, RL-SPAACE is agnostic to the level of control in a nested feedback control loop. For this case, an outer voltage loop is added to the inner current loop of an IBR to control its terminal voltage. In the test system in Fig.~\ref{fig:CIGRE-14}, the switch~S4 is opened to disconnect the test system from the grid. A single IBR is connected to the test system by closing switches S1, S2, or S3. Since only one IBR is utilized to energize the test system, preset frequency and voltage references are provided to this IBR. The outer voltage loop PI controller coefficients are set to $k_{PD}$=0.01 and $k_{ID}$=0.0005 for the d-axis controller, and $k_{PQ}$=0.01 and $k_{IQ}$=0.0003 for the q-axis controller. The inner current loop PI controller coefficients are set to $k_{P}$=1.0 and $k_{I}$=0.003 for both the d- and q-axis controllers. The value of the capacitor in the LCL filter of the IBR is increased to exaggerate the level of overshoot in the base case and verify the utility of RL-SPAACE in controlling this overshoot. The d-axis component of the IBR terminal voltage $v_d$ is set to 0.1 pu and q-axis component of the IBR terminal voltage $v_q$ is set to 0.0 pu. In each scenario, a step change is applied to the q-axis component of the IBR terminal voltage with an upper limit of 1.1 pu, while the respective d-axis component is held at 0.1 pu. Algorithm~1 is utilized in training the controller.

After sufficient training, the same test system is utilized to test the performance of learned control strategy in islanded mode. For this case, the switch~S2 is closed, and the other switches~S1, S3, and S4 are opened. At the start of the simulation, the IBR2 terminal voltage references $v_d$ and $v_q$ are set to 0.1 pu and 0.0 pu, respectively. At a time $t=\SI{0.2}{s}$, $v_q$ is increased in step to 0.9 pu, while $v_d$ is unchanged. RL-SPAACE communicates with the test system every \SI{100}{\micro\s}.

\begin{figure}[!t]
    \centering
    \subfloat[]{\includegraphics[width=0.85\columnwidth, trim=8 8 6 8,clip]{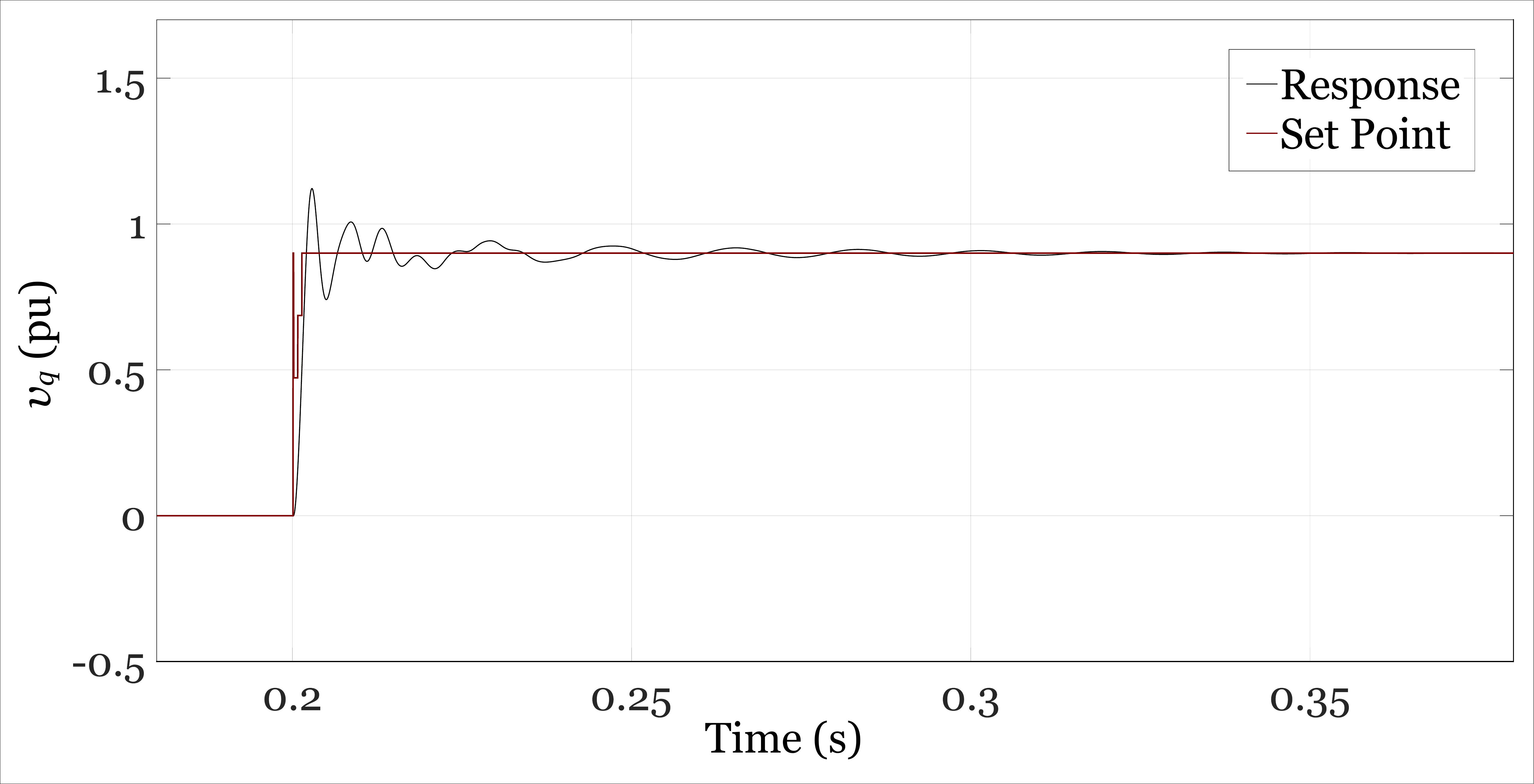}\label{fig:results5a}}\\[-0.2ex]
    \subfloat[]{\includegraphics[width=0.85\columnwidth, trim=8 8 6 8,clip]{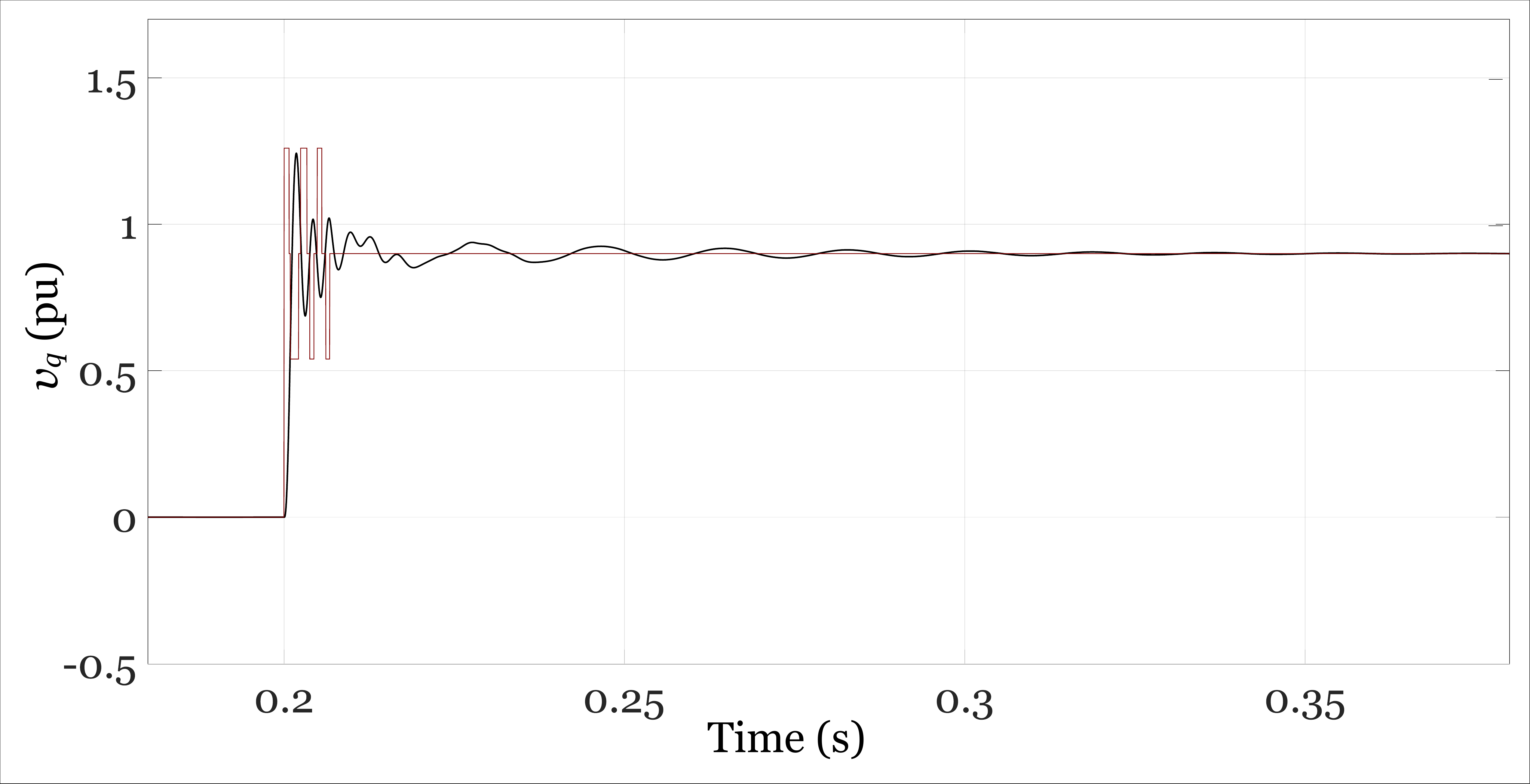}\label{fig:result5b}}\\[-0.2ex]
    \subfloat[]{\includegraphics[width=0.85\columnwidth, trim=8 8 6 8,clip]{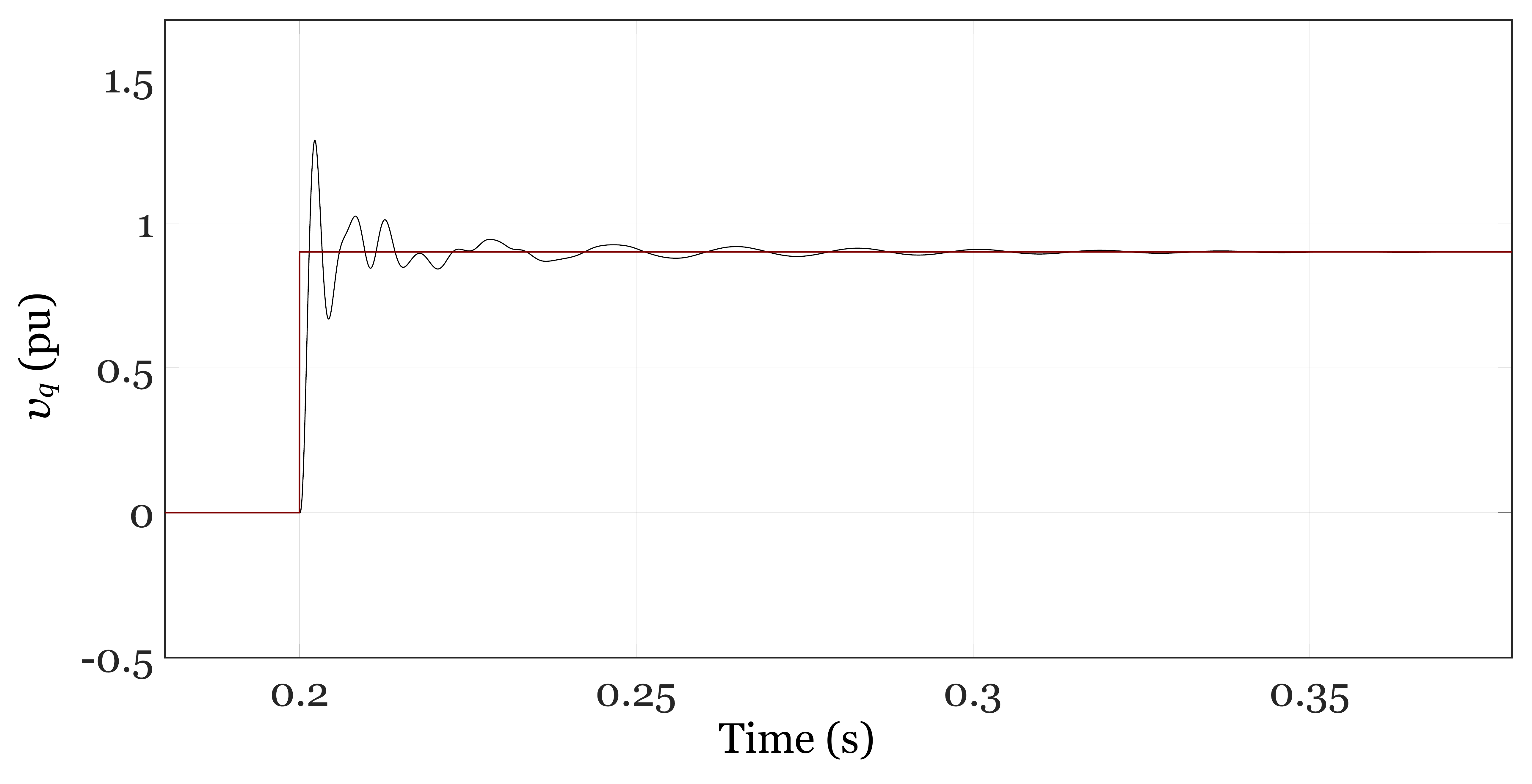}\label{fig:result5c}}
\caption{Case 1: Step change in IBR2 voltage from 0.0 pu to 0.9 pu (a) with RL-SPAACE, (b) with SPAACE, and (c) without any supplementary action.}
\label{fig:results5}
\end{figure}

Similar to previous case, the performance of RL-SPAACE is compared with SPAACE utilizing linear prediction and without any supplementary control. For SPAACE, the value of $m$ is set to 0.4 pu and $t_{pred}$ is set to \SI{500}{\micro\s}. The minimum and maximum values of $v_q$ are set to 0.8 pu and 1.2 pu, respectively. SPAACE communicates with the test system every \SI{10}{\micro\s} Fig.~\ref{fig:results5} shows the results from this case. The results show that RL-SPAACE results in 24.4\% overshoot in response, while SPAACE and the case without any supplementary control result in 37.7\% and 42.2\% overshoot, respectively. The results from this case prove that RL-SPAACE using Algorithm~1 could be utilized in improving transient performance of inverters forming a grid.

\section{Conclusion}\label{sec:conclusion}
In this paper, RL-SPAACE, a control strategy to limit transients in IBRs connected to microgrids is proposed. This strategy utilizes RL to issue adaptive set point changes by monitoring the current state and trajectory of the response variable. The strategy is implemented on an example microgrid test system to control transients resulting from both planned set point changes and unplanned contingencies in IBRs. A significant improvement in transient performance of IBR is reported through utilization of the proposed control strategy.

\bibliographystyle{IEEEtran}
\bibliography{RLS}

\vspace{12pt}

\end{document}